\documentclass{emulateapj}
\usepackage{times}
\usepackage{amsmath}
\usepackage{natbib}
\usepackage{multirow}

\begin{document}

\shorttitle{Unmasking the AGN in J2310-437}
\shortauthors{Bliss et al.}

\title{Unmasking the Active Galactic Nucleus in PKS J2310-437 }

\author{A.F.Bliss, D.M.Worrall, M Birkinshaw}
\affil{H H Wills Physics Laboratory, Tyndall Avenue, Bristol, BS8 1TL, UK}

\author{S.S.Murray and H.Tananbaum}
\affil{Harvard-Smithsonian Center for Astrophysics, Cambridge, MA 02138}

\begin{abstract}

PKS J2310-437 is an AGN with bright X-ray emission relative to
its weak radio emission and optical continuum. It is believed that its jet lies far enough from the line of sight that it is not highly relativistically beamed.  It thus provides an extreme test of AGN models.  We present new
observations aimed at refining the measurement of the source's
properties. In optical photometry with the NTT we measure a central
excess with relatively steep spectrum lying above the bright
elliptical galaxy emission, and we associate the excess wholly or in
part with the AGN.  A new full-track radio observation with the ATCA
finds that the core 8.64~GHz emission has varied by about 20 per cent over
38 months, and improves the mapping of the weak jet.  With {\it
Chandra\/} we measure a well-constrained power-law spectral index for
the X-ray core, uncontaminated by extended emission from the
cluster environment, with a negligible level of intrinsic absorption.
Weak X-ray emission from the resolved radio jet is also measured. Our analysis
suggests that the optical continuum in this radio galaxy has
varied by at least a factor of four over a timescale of about two
years, something that should be testable with further observations. We conclude that the most likely explanation for the bright central X-ray emission is synchrotron radiation from high-energy electrons.

\end{abstract}


\section{Introduction}

PKS J2310-437 is hosted by an elliptical galaxy at redshift $z=0.0886$, at the
centre of a cluster of Abell richness 0 \citep{ttr95}.  It is
interesting and unusual in that it appears as a luminous ($\sim10^{44}
{\rm ergs\ s} ^{-1}$) X-ray source, associated with a significant
radio source \citep{Tetal, Wetal}, but with low optical luminosity and
the characteristics of a typical elliptical galaxy.  It shows no
optical line emission as could be expected from an AGN, or UV
continuum emission as from a BL Lac nucleus.  This is not believed to
be due to extinction as ROSAT X-ray data demonstrate a soft spectrum
showing no excess absorption by gas in the J2310-437 galaxy
\citep{Tetal}, and it would be hard to argue for a dust/gas torus that
absorbs the AGN optical light without diminishing the soft X-ray emission.

Any optical AGN emission must be weak.
\citet{Wetal}
give an upper limit of 32 $\mu$Jy at 4400\AA, based on the size of
the CaII break in a spectrum obtained in 1996, but suggested
that the actual value lies close to
this limit.  Evidence for the presence of a weak AGN optical continuum
is presented by \citet{Cetal} who find that the size of the CaII break in J2310-437 as measured
through a narrow spectral slit to be 38 $\pm4\%$, which lies between the values
expected of a BL lac object ($\leq25\%$) and that of a typical
elliptical galaxy ($\approx50\%$).  They measured a larger break when
using a wider spectral slit that would incorporate the same AGN
emission diluted by more starlight.  However,
no absolute flux calibration was available to put qualitative evidence
for AGN emission onto a more quantitative footing.

The radio structure is interesting, showing a one-sided jet that
extends about 10 arcsec to the south-east of the core, embedded in a
large-scale double plume \citep{Wetal}.  That relatively large radio
lobes can be
seen on both sides implies that J2310-437 is seen less `end-on' than expected of a BL Lac object. \citet{Wetal} suggest
that the jet is $\gtrsim30^{\circ}$ to the line of sight based on
a core prominence that is lower than expected for a BL Lac object.

Objects that show extreme characteristics in their multifrequency
properties are likely to challenge most strongly our understanding of
their emission mechanisms.  In this paper we present new observations
aimed at testing how well the extreme properties of J2310-437 might be
fitted within the framework of emission models discussed for other
AGN.

The redshift of J2310-437 is 0.0886.  In this paper we adopt values
for the cosmological parameters of $H_0 = 70$~km s$^{-1}$ Mpc$^{-1}$,
$\Omega_{\rm {m0}} = 0.3$, and $\Omega_{\Lambda 0} = 0.7$.  Thus
1~arcsec corresponds to a projected distance of 1.66~kpc at the
source.

\section{Observations}\label{sec:observations}

\subsection{Observational objectives}\label{sec:objectives}
We observed J2310-437 with the SUSI2 instrument on the European
Southern Observatory New Technology Telescope (ESO NTT) with the
intent of exploiting its excellent seeing to detect the
optical core of the galaxy and image the jet.  We also
observed J2310-437 with {\it Chandra\/} to obtain
high-resolution spatial and spectral data so that the core AGN emission spectrum could be resolved from the galaxy and cluster, and also to
search for an X-ray jet corresponding to the radio jet.  Finally, new
radio data were obtained using the Australian
Telescope Compact Array (ACTA).  Two 12-hour tracks, 51 months apart,
have improved on the earlier results reported by
\citet{Wetal}.\\*


\subsection{Optical Observations}\label{sec:opticalobs}

Data in the UBVR- and I-bands were collected with the ESO 3.5m NTT
equipped with the Superb Seeing Imager - 2 instrument (SUSI2) on the
nights of 2004 August 7, and 2004 September 13 and 22.  The SUSI2 CCD
camera provides a field of view 5.5 x 5.5 arcmin$^2$, with a scale of 0.16 arcsec pixel$^{-1}$ (after 2x2 binning).  Total
integration times, central wavelengths and average seeing are given in Table
\ref{tab:filters}.

\begin{table} 
\caption{SUSI2 Exposures} \label{tab:filters}
\centering
\begin{tabular} {c c c c c c}
\hline \hline
Band & U & B & V & R & I \\
\cline{1-6}
Central $\lambda$ (nm) & 357.08 & 421.17 & 544.17 & 641.58 & 794.96\\
Integration time (s) & $10\times 240$ & $7\times 60$  & $6\times 60$  &  $6\times 60$ & $7\times 120$ \\
FWHM (arcsec) & 1.05 & 0.64 & 0.63 & 1.30 & 0.40\\
\hline
\end{tabular}
\label{table:filters}
\end{table}

\subsubsection{Optical Data reduction}\label{sec:opticalreds}

The raw data were processed using IRAF to bias-subtract, flat-field
and apply gradient corrections.  Dome flats were used for
flat-fielding.  The I-band images contained strong fringing which was
removed by creating and subtracting a master fringe mask.  Any
remaining gradient was removed by subtracting a fitted background
using IRAF.  The images were then flipped along a vertical axis to
bring them to sky orientation.  A WCS was added to the images
using an R-band optical image with attached world co-ordinate system
that was previously acquired from the CTIO telescope. A median average was used to combine the data to create one mosaic for each band.This proved effective at removing cosmic ray hits. The number of overlapping
frames used to create the mosaics were 10, 7, 6, 6 and 7 in U, B, V, R
and I respectively.  The seeing was particularly bad for the R-band
data (Table~\ref{tab:filters}), and so data in this filter are
excluded from our subsequent analysis.  Flux calibration was carried
out using the photometry for standard stars PG1525A and PG1525B given
in the Landolt Equatorial Standards table \citep{landolt}.
.

\subsubsection{Radial Profile Extraction}\label{sec:opticalradial}

Radial profiles in each optical band were extracted from the
background-subtracted mosaics, and the count rates in each annulus
were measured using IRAF.  The count-rate error contributed by
background subtraction was calculated by fitting a Gaussian to the
histogram of pixel counts in the non-source regions of the
background-subtracted images. The annular errors were calculated based
on Poisson statistics and combined in quadrature with the background
errors.

\subsubsection{Galactic Modelling}\label{sec:modelling}
To measure the flux density from the AGN the stellar component of the
galaxy needed to be removed.  This was carried out using GALFIT
\citep{Peng} which is an algorithm that fits 2-D parameterized models
of galaxies and/or point sources to images.  As J2310-437 lies in an
elliptical galaxy the model used was a S\'{e}rsic profile \citep{Sersic} of
the form

\begin{equation*}
 \Sigma(r)=\Sigma_{e} \exp \biggl[-\kappa \biggl(\Bigl(\frac{r}{r_e}\Bigr)^{\frac{1}{n}}-1 \biggr)\biggr]
\end{equation*}

\noindent
where $r_{e}$ is the effective radius of the galaxy and $\Sigma_{e}$
is the surface brightness at $r_{e}$.  Cuspy galaxies have low values
of the S\'{e}rsic index, n, and flatter-profile galaxies have higher n: n
= 4 is the `classic' de Vaucouleurs profile.  $\kappa$ is coupled to n
such that half the total flux is within $r_{e}$.  First, all objects
in the mosaiced image were masked and a sky level found. This was
close to zero in the background-subtracted mosaics.  A point spread
function (PSF) for each band was created from stars in the field,
using the DAOPHOT package in IRAF. Table~\ref{tab:filters} gives the
FWHM of this PSF for each band. Then a central area twice the radius of the
FWHM was masked on the galaxy, to mask the AGN contribution, and a
S\'{e}rsic profile convolved with the PSF was fitted to the data.  The
parameters from the S\'{e}rsic model were then fixed and extrapolated into the central region. This model was subtracted leaving a residual image. The central flux excess was then measured from this residual image.
The uncertainties in the flux excess were found by using combinations
of S\'{e}rsic index, magnitude and effective radius (within uncertainties)
that gave minimum and maximum contributions to the measured excess
emission.  These were added linearly with the statistical errors to
give a total error on the excess (AGN) emission.

\subsection{X-ray Observations}\label{sec:x-rayobs}

J2310-437 was included in the Chandra AO3 HRC GTO target list, and the source
was observed on 2002 July 26 with the S3 back-illuminated CCD of the
{\it Chandra\/} ACIS-S for $\sim$30~ks.  The exposure was designed to take
advantage of {\it Chandra}'s excellent spatial resolution to search
for X-ray emission corresponding to the radio jet and to measure the
spectrum of the core uncontaminated by emission from the cluster gas
seen in {\it ROSAT\/} data. ACIS-S was used in 1/8 window mode to
reduce the effect of pile up since the AGN is a strong
point-like X-ray emitter. The pipeline data products were re-analyzed using
CIAO~3.4 to apply recent calibrations, including corrections for the build up of contaminants on the ACIS
that result in losses at low energy. The pixel randomization routinely
applied to the ACIS data was turned off.  We used a grade selection of
0,2,3,4,6.  Since the data were taken in the very faint (VF) mode, we
used VF screening to help clean the data and improve the signal to
noise when searching for low surface brightness features associated
with the radio jet.  This procedure can reject valid events in regions
where the count rate is high (i.e., the core of the AGN). Thus these
data were restored ($\sim 3.6$\% of the counts) for our spectral analysis
of the core. The ACIS readout streak that results from ``out of time''
events from the bright central core was removed for
morphological analysis of the image.  The readout streak is at a
position angle of $\sim 110^\circ$, and so does not contaminate the
X-ray emission in the direction of the radio jet (see
Fig.~\ref{fig:xband}).  There were no periods of high background and
the final calibrated data have an exposure time of 26.75~ks.

\subsection{Radio Observations}\label{sec:radioobs}

New radio observations of J2310-437 were made at C and X bands using
the ATCA in 6D array with 6 antennas on 2000 April 4-5. Data were taken in full polarization over
128~MHz bands centered on 4.80 and 8.64~GHz, with flux density and
bandpass calibrations based on PKS~1934-638, with assumed flux
densities of 5.83 and 2.84~Jy, respectively. Calibration and imaging
of the data followed standard techniques, with several
self-calibration cycles, before the data were combined with the
earlier (1997 January 29-30) data used in \citet{Wetal}.

\begin{figure}
\plotone{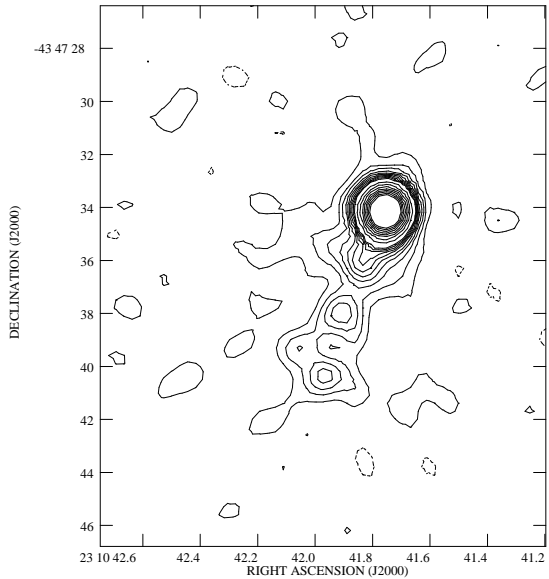}
\caption{\footnotesize 8.64-GHz radio image of J2310-437 made from two full tracks with the ATCA.  A strongly one-sided jet emerges from the
flat-spectrum core. The peak flux density is 18.9~mJy beam$^{-1}$.  Contours are at $0.2 \times (-1, 1, 2, 3, 4, 5, 6, 7, 8, 9, 10, 15, 20, 25, 30, 35,
40, 45, 50)$ mJy beam$^{-1}$, with the negative contour dashed. The restoring beam is a $1.2$ arcsec FWHM circular Gaussian.\label{fig:xband}}
\end{figure}

The new X-band image of J2310-437 based on both datasets is shown in
Figure \ref{fig:xband}, and provides a better representation of the
centre of the source than in the earlier image \citep{Wetal}, in particular in
providing better information on the kpc-scale structure of the jet because of the improved weather and the presence of shorter antenna-antenna spacings in the array configuration in the 2000 observing session.  .
The core position is measured
to be at $\alpha = \rm 23^h10^m41^s\llap{.}756 \pm 0^s\llap{.}007$,$\delta =
-43^\circ47^\prime34^{\prime\prime}\llap{.}2 \pm
0^{\prime\prime}\llap{.}1$, a slight refinement from the position given in \citet{Wetal}. 
Fits to the images from the two epochs, taking account of the
different data samplings by fitting the core as above extended jet
emission using the AIPS task IMFIT, find that the core flux density
has decreased from $20.9 \pm 0.3$~mJy to $16.2 \pm 0.3$~mJy at X~band
over 38~months, while the flux density has increased from $20.6 \pm
0.5$ to $24.4 \pm 0.5$~mJy at C~band. This indicates both luminosity
and spectral variability of the core.

The resolved (non-core) jet emission has flux densities of $16 \pm
2$~mJy at 8.64~GHz and $20 \pm 2$~mJy at 4.8~GHz within 10 arcsec of
the core, corresponding to a spectral index of $0.4 \pm 0.3$, and
consistent with the typical spectral index $\sim 0.6$ found in
kpc-scale radio jets.

\begin{figure}
\plotone{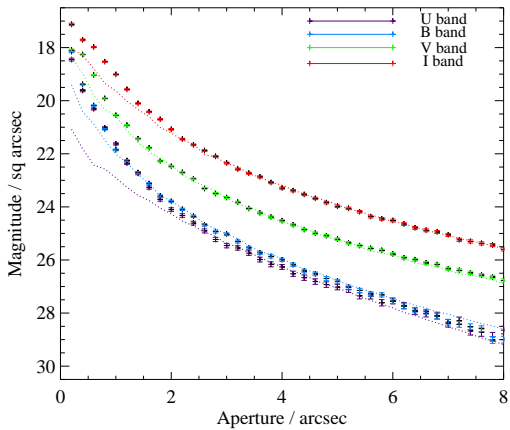}
\caption{\footnotesize Surface Brightness Profile with the Sersic models as determined by Galfit shown as dotted lines. Errors are 1$\sigma$ calculated as in \S\ref{sec:opticalradial}. \label{fig:surf_model}}
\end{figure}

\section{Results}\label{sec:results}

\subsection{Optical}\label{sec:resultsoptical}

Figure \ref{fig:surf_model} shows the surface brightness profile of
each optical band with the S\'{e}rsic model-profile, extrapolated into the
center, overlaid.  The data show an excess over the S\'{e}rsic
profile within $\sim 2''$ of the
nucleus.  The best-fit parameters
for the S\'{e}rsic models are given in Table \ref{tab:sersic},
including total magnitudes integrated out to infinity.


\begin{table*}
\caption{S\'{e}rsic Parameters} \label{tab:sersic}
\centering
\begin{tabular}{c c c c c}
\hline \hline
Band & U & B & V & I\\
\cline{1-5}
S\'{e}rsic Index, n & 1.25$\pm$0.08 & 2.59$\pm$0.05 & 3.15$\pm$0.04 & 2.52$\pm$0.02\\
Effective Radius, $r_{e}$ (arcsec) & 3.29$\pm$0.08 & 3.06$\pm$0.03 & 4.56$\pm$0.04 & 4.58$\pm$0.02\\
Integrated Magnitude & 18.54$\pm$0.03 & 17.59$\pm$0.01 & 15.92$\pm$0.01 & 14.65$\pm$0.01\\
\hline
\end{tabular}
\label{tab:sersic}
\end{table*}

\begin{figure*}
 \plotone{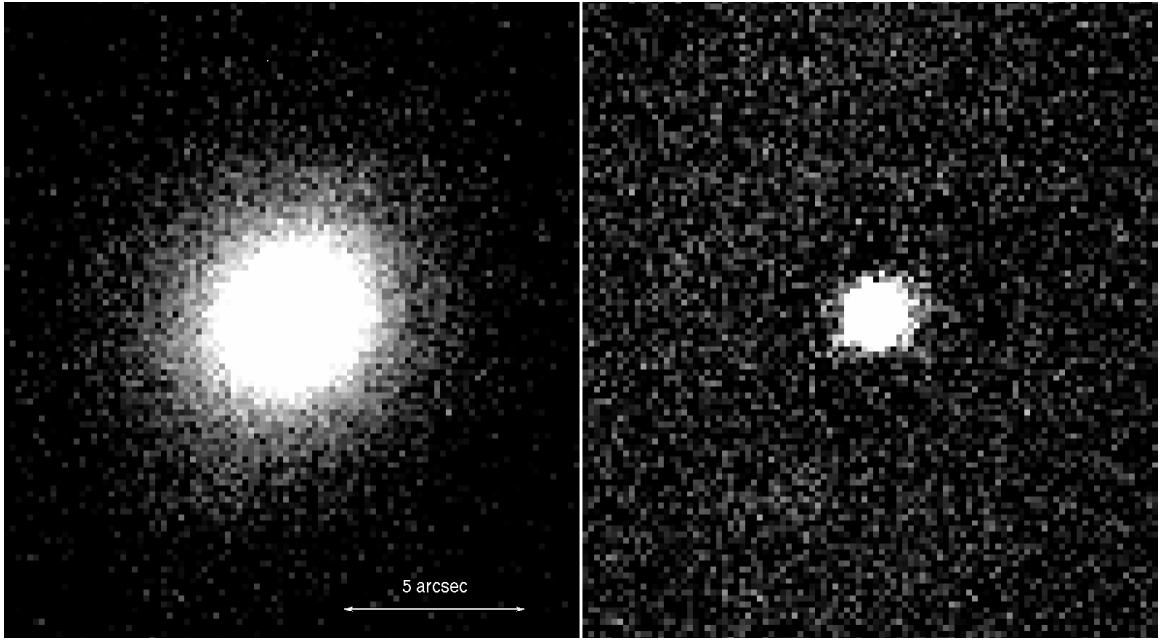}
\caption{\footnotesize Left panel shows B-band image of J2310-437 in
its host galaxy, and the right panel shows remaining excess after galactic
modelling and subtraction. \label{fig:bband}}
\end{figure*}

Figure \ref{fig:bband} shows the J2310-437 host galaxy in the B filter
on the left, and on the right the remaining excess after subtracting
the galactic model fitted by GALFIT. The residuals were free of
obvious structure related to the galaxy morphology, as can be seen in
Figure \ref{fig:bband}.
The values of residual (AGN) flux density are given in Table \ref{tab:flux density}.

\begin{table}
\caption{AGN Flux Density} \label{tab:flux density}
\centering
\begin{tabular} {c c c c}
\hline \hline
Band & Frequency (Hz) & Flux Density ($\mu$Jy) & Error ($\mu$Jy)\\
\cline{1-4}
\multirow{2}{*}{U} & \multirow{2}{*}{$8.40\times10^{14}$} & \multirow{2}{*}{41.12} & \scriptsize{$+0.89$}\\
& & & \scriptsize{$-0.90$}\\
\cline{1-4}
\multirow{2}{*}{B} & \multirow{2}{*}{$7.12\times10^{14}$} & \multirow{2}{*}{59.48} & \scriptsize{$+1.77$}\\
& & & \scriptsize{$-2.21$}\\
\cline{1-4}
\multirow{2}{*}{V} & \multirow{2}{*}{$5.51\times10^{14}$} & \multirow{2}{*}{139.16} & \scriptsize{$+4.98$}\\
& & & \scriptsize{$-4.75$}\\
\cline{1-4}
\multirow{2}{*}{I} & \multirow{2}{*}{$3.77\times10^{14}$} & \multirow{2}{*}{242.63} & \scriptsize{$+8.17$}\\
& & & \scriptsize{$-8.66$}\\
\hline \tabularnewline
\end{tabular}
\label{tab:flux density}
\end{table}

\subsection{X-ray}\label{sec:x-ray results}

\subsubsection{Spatial Structure}\label{sec:X-rayspatial}

\begin{figure*}

\plottwo{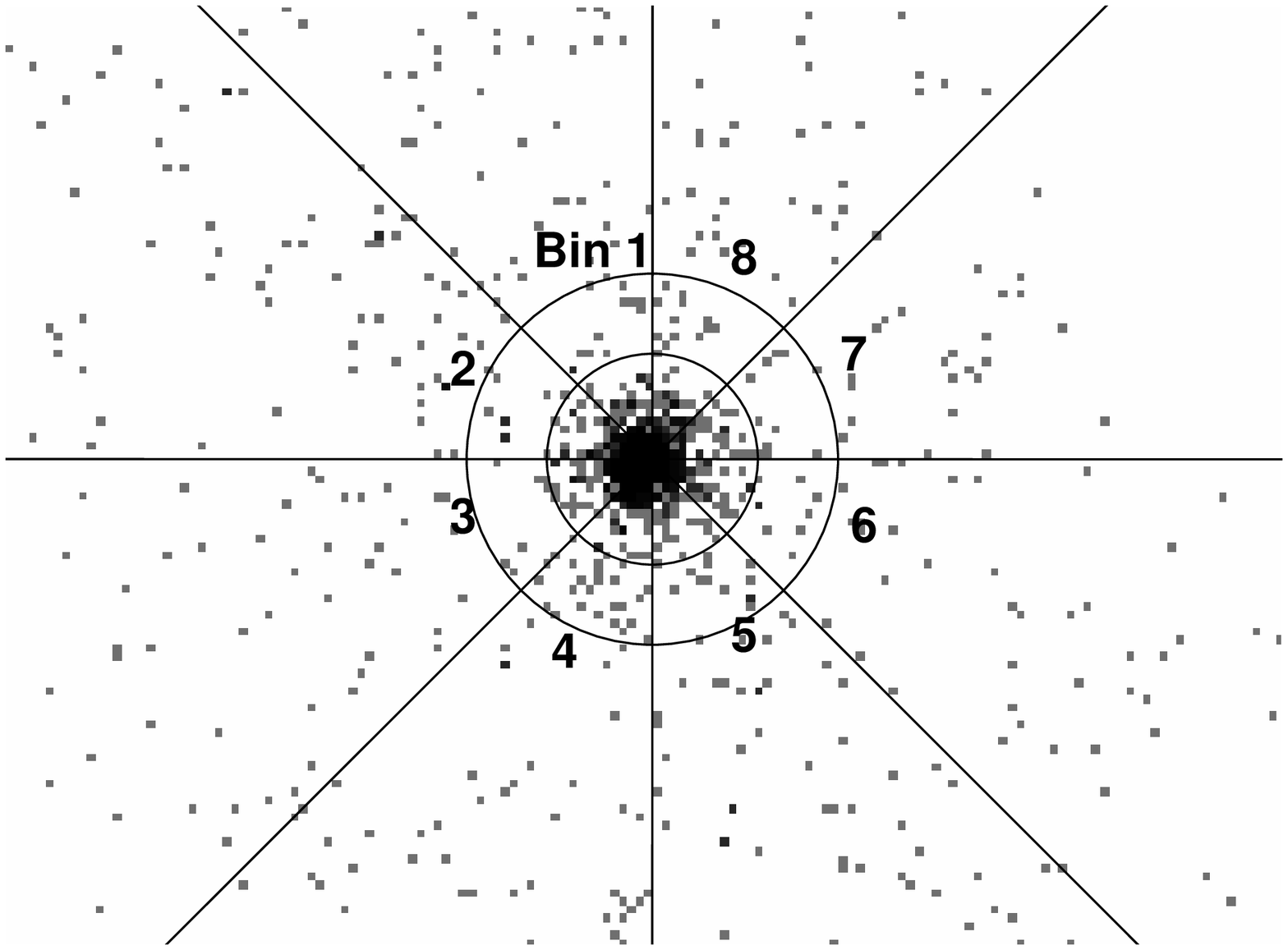}{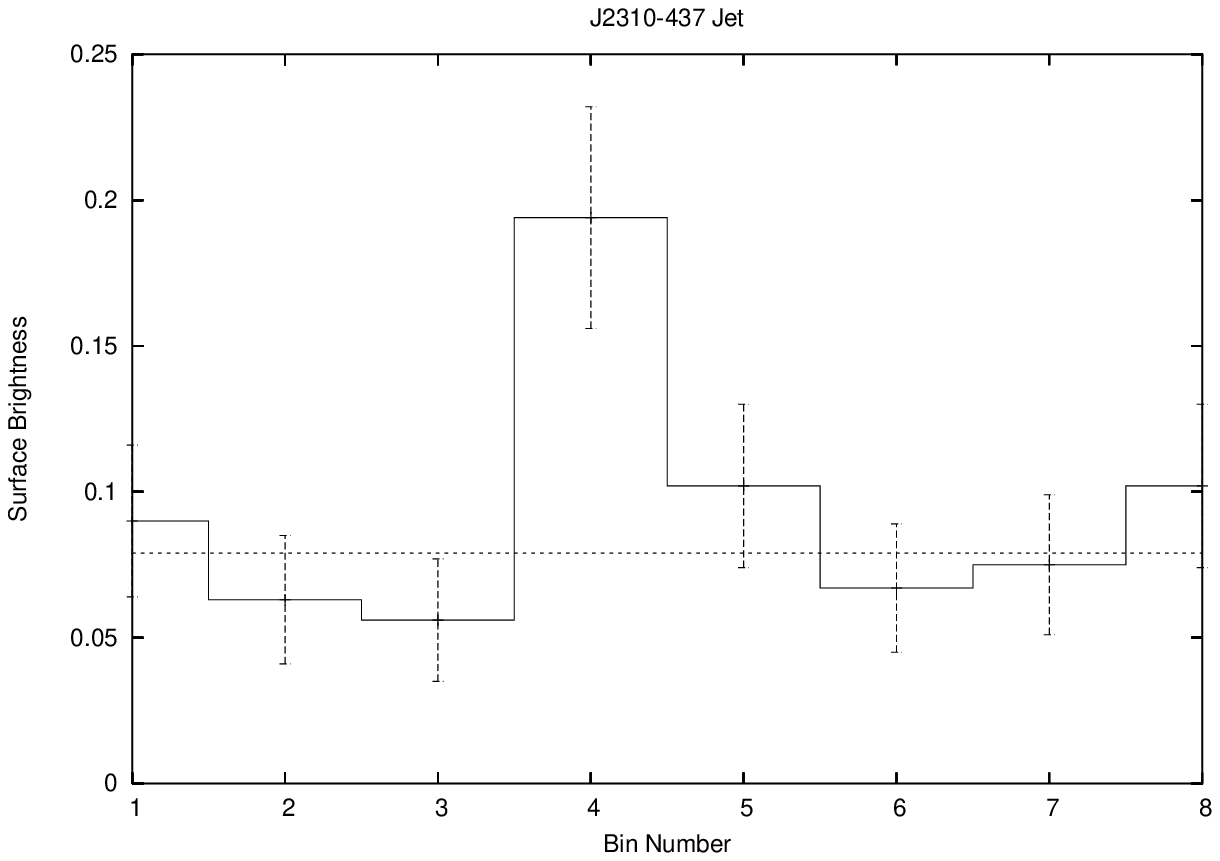}
\caption{\footnotesize Left panel shows the cleaned {\it Chandra\/} X-ray image with the
annular region from which the azimuthal count data were extracted. The
inner radius is 12.5 pixels and the outer radius is 22 pixels. The
annulus is divided into 8 azimuthal sectors with the bins labelled (see Table \ref{tab:azimuth}). On
the right, the azimuthal bin data are plotted as counts per pixel. The surface brightness
peak in bin 4 coincides with the orientation of the inner radio
jet. Error bars are 1$\sigma$.\label{fig:azimuth}}
\end{figure*}

Visual inspection of the image showed evidence for weak X-ray emission
at the position angle of the radio jet.  To investigate this further,
the cleaned data (0.3-7~keV) were summed in 8 azimuthal
source-centered bins in an annulus with radii of 12.5 and 22 ACIS
pixels (6.15 - 10.83 arc seconds) as shown in
Figure~\ref{fig:azimuth}. The total counts in each azimuthal bin (bins
are numbered sequentially moving counter clockwise in the figure) are
given in Table \ref {tab:azimuth}. In the analysis, events were
assigned as being completely in or out of a bin (no fractional counts)
and the integer number of image pixels in each bin was used to find the surface brightness, $\Sigma_X$.  Bin 4 is co-located with the outer part of the
radio jet shown in Figure~\ref{fig:xband}.

\begin{table}

\caption{Azimuthal Projection Data}\label{tab:azimuth}

\begin{centering}{\footnotesize }\begin{tabular}{|c|c|c|c|c|c|}
\hline
{\footnotesize Bin}&
{\footnotesize Counts}&
{\footnotesize Error}&
{\footnotesize Pixels}&
{\footnotesize $\Sigma_X$ counts/pixel}&
{\footnotesize $\Sigma_X$ Error}\tabularnewline
\hline
\hline
{\footnotesize 1}&
{\footnotesize 12}&
{\footnotesize 3.5}&
{\footnotesize 134}&
{\footnotesize 0.090}&
{\footnotesize 0.026}\tabularnewline
\hline
{\footnotesize 2}&
{\footnotesize 8}&
{\footnotesize 2.8}&
{\footnotesize 126}&
{\footnotesize 0.063}&
{\footnotesize 0.022}\tabularnewline
\hline
{\footnotesize 3}&
{\footnotesize 7}&
{\footnotesize 2.6}&
{\footnotesize 126}&
{\footnotesize 0.056}&
{\footnotesize 0.021}\tabularnewline
\hline
{\footnotesize 4}&
{\footnotesize 26}&
{\footnotesize 5.1}&
{\footnotesize 134}&
{\footnotesize 0.194}&
{\footnotesize 0.038}\tabularnewline
\hline
{\footnotesize 5}&
{\footnotesize 13}&
{\footnotesize 3.6}&
{\footnotesize 127}&
{\footnotesize 0.102}&
{\footnotesize 0.028}\tabularnewline
\hline
{\footnotesize 6}&
{\footnotesize 9}&
{\footnotesize 3.0}&
{\footnotesize 134}&
{\footnotesize 0.067}&
{\footnotesize 0.022}\tabularnewline
\hline
{\footnotesize 7}&
{\footnotesize 10}&
{\footnotesize 3.2}&
{\footnotesize 134}&
{\footnotesize 0.075}&
{\footnotesize 0.024}\tabularnewline
\hline
{\footnotesize 8}&
{\footnotesize 13}&
{\footnotesize 3.6}&
{\footnotesize 127}&
{\footnotesize 0.102}&
{\footnotesize 0.028}\tabularnewline
\hline
\end{tabular}\par\end{centering}
\end{table}

The mean number of counts in each azimuthal bin, excluding bin 4, is
10.3. This is significantly higher than the equivalent background rate
far from the core source ($\sim$4) and predominantly arises from the low level of
scattered core events by the {\it Chandra\/} telescope.  In bin 4 there are
26 counts.  The Poisson probability of 26 or more counts given a mean
of 10.3 is $2.87\times10^{-5}$, which is equivalent to about a
$4\sigma$ significance for existence in a Gaussian sense. (Note if the
overall azimuthal mean of 12.3 counts bin$^{-1}$ is taken instead,
then the Poisson probability of 26 or more counts in bin is
$4.2\times10^{-4}$, equivalent to about a $3.5\sigma$ significance for
existence in the Gaussian sense.) Taking into account the prior
knowledge of the radio jet location, then the coincidence of the X-ray
excess counts becomes more significant.  The jet is detectable in the
X-ray only because of the excellent performance of the {\it Chandra\/}
X-ray telescope, both in terms of the central point spread function,
and the very low scatter. The soft spectrum of the core (see
\S\ref{sec:core}) is helpful since soft X-rays are scattered less by
the telescope.

In a 4-pixel (1.97-arcsec) radius circular region centered on the core
the total number of counts is $\sim 12,000$. Therefore the jet feature
located between 6.15 and 10.83 arcsec from the core contains only
$\sim 0.2$\% of the flux of the core. In the radio, the corresponding region contains $\sim 15$\% and $\sim 10$\% of the core flux density at 4.8 and 8.64~GHz,
respectively.  We did not detect the resolved jet in our optical observations and the high level of galaxy light means there are no useful limits to its flux in the NTT bands.

\begin{figure}
\plotone{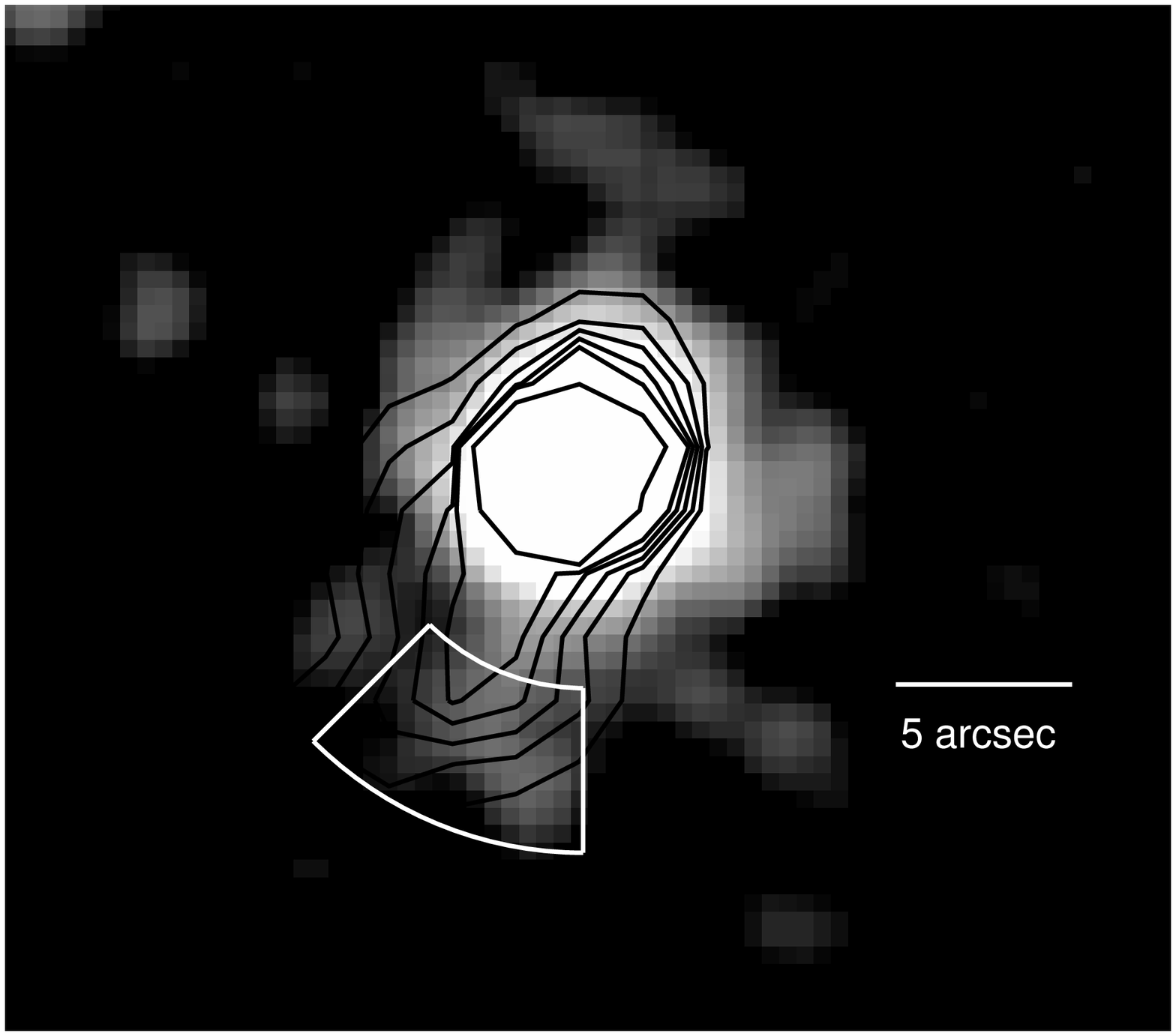}
\caption{\footnotesize Smoothed {\it Chandra\/} X-ray image of
J2310-437 with 4.8~GHz radio contours (black) overlaid.  The peak
radio flux density is 24.4 mJy beam$^{-1}$.  Contours are at $0.2
\times (1,2,3,4,5,10)$ mJy beam$^{-1}$ and the beam size is $2 \times 2$
arcsec$^2$.  The region in white corresponds to bin 4 of
Figure~\ref{fig:azimuth}. \label{fig:x-rayimage}}

\end{figure}

Figure~\ref{fig:x-rayimage} shows a smoothed X-ray image where a
Gaussian of $\sigma = 1.5$ pixels (0.75 arcsec) is convolved with the
data.  The color map is logarithmic and stretched to emphasize the low
count-rate features. The black contours are of the 4.8~GHz radio
emission from combining the full tracks with the ATCA from 1997 and
2000, and the white region corresponds to bin 4 of
Figure~\ref{fig:azimuth}.

\subsubsection{Core Spectrum}\label{sec:core}

The X-ray spectrum of the core was extracted from a source-centered
circle of radius 4.3 arcsec, with background from a source-centered
annulus of radii 6.3 and 15 arcsec. Using {\sc xspec} the data fitted
well a power-law model with an absorption column density consistent
with that for the line of sight through our Galaxy of $1.57 \times
10^{20}$ cm$^{-2}$.  The fit gave $\chi^2 = 206$ for 176 degrees of
freedom for a fit to the counts over 0.4-7~keV ($\sim 12000$ nets
counts) grouped to a minimum of 25 counts per bin.  Best-fit parameter
values with $1\sigma$ errors for 2 interesting parameters are
$\alpha_{\rm x} = 0.91\pm 0.05$, $N_{\rm H} = (1.2 \pm 1.0) \times
10^{20}$ cm$^{-2}$.  There was no significant improvement of fit if a
thermal component was added to the model.  This is not surprising: although
X-ray emission from cluster gas was measured in {\it ROSAT\/}
\citep{Tetal, Wetal}, the surface brightness is low, estimated at
$\sim 0.2$ counts arcsec$^{-2}$ (or about 3 total counts) under the
core in the {\it Chandra\/} exposure.

\begin{figure}
\includegraphics[height=0.9\columnwidth, clip=true, angle=270]{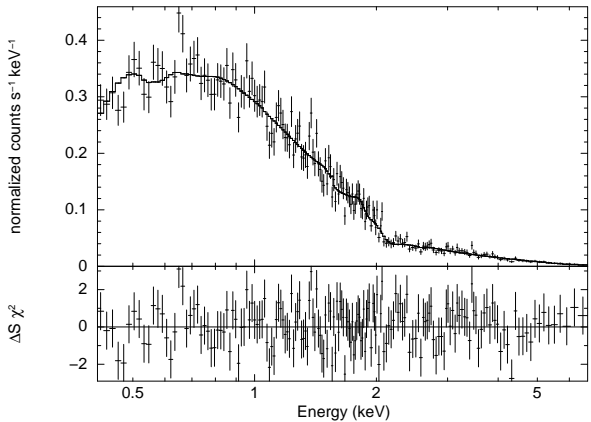}
\caption{\footnotesize X-ray spectrum of the core fitted to a pileup
power-law model with $N_{\rm H}=1.57 \times 10^{20}$ cm$^{-2}$,
$\alpha_{\rm x} = 0.98$, and normalization corresponding to a 1-keV
flux density of $0.435\mu$Jy. The lower panel shows residuals (data - model) expressed as their contribution to $\chi^2$. $\chi^2_{\rm min} = 203$ for 175
degrees of freedom.\label{fig:xrayspectrum}}
\end{figure}
\begin{figure}
\includegraphics[height=0.6\columnwidth,clip=true,angle=270]{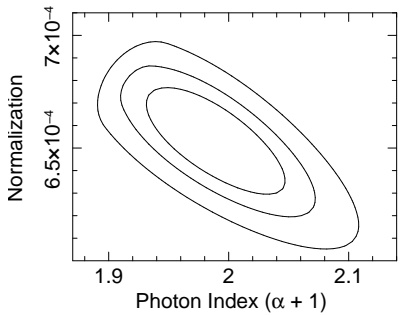}
\caption{\footnotesize $\chi^2$ contours corresponding to $1\sigma$,
90\%, 99\% uncertainties for 2 interesting parameters for the X-ray
spectral fit of the AGN core to a pileup power-law model with $N_{\rm
H}$ fixed to the Galactic value.  The normalization is in units of keV
cm$^{-2}$ s$^{-1}$ keV$^{-1}$ at 1~keV.\label{fig:xraycontour}}
\end{figure}

The high count-rate from the core means that there are expected to be
instances when two photons arrive within the readout time of the CCDs
and are measured as one event of higher energy.  We had alleviated the
affect, known as pileup, by reducing the frame time to 0.4~s through
reading out only 1/8 of the S3 chip.  The {\sc xspec} and {\sc sherpa}
spectral-fitting programmes each incorporate models which seek to take
into account pileup.  As expected, when we allowed for this the
spectrum became slightly steeper and the 1 keV normalization was
slightly higher, but still a single-power law with a low Hydrogen
column density was the preferred underlying model.  The level of
pile-up was indicated as $\sim 5$\%, and in both cases there was a
decrease in $\chi^2$ of roughly 3.  Results for $N_{\rm H}$ were
similar to the fits which did not allow for pileup, and so we froze
$N_{\rm H}$ at the Galactic value.  The spectrum is shown in
Figure~\ref{fig:xrayspectrum} and uncertainty contours in spectral
index and normalization are given in Figure~\ref{fig:xraycontour}.
Best-fit parameter values with $1\sigma$ errors for 2 interesting
parameters are now $\alpha_{\rm x} = 0.98^{+0.07}_{-0.05}$ with the
normalization corresponding to a 1-keV flux density of $0.435 \pm
0.015\mu$Jy.  The 1-keV flux density is in good agreement with results
using the {\it ROSAT\/} PSPC and HRI.  The spectral index is much
better constrained because the PSPC was the only {\it ROSAT\/}
detector with spectral information, and its relatively large PSF mixed
$\sim 20$\% X-ray emission from cluster gas with the emission from the
AGN.

While the pileup model gives a value for $N_{\rm H}$ that is
consistent with the Galactic value, our fits allow an additional small
amount of column density intrinsic to the source, $N_{\rm H_{int}}$.
The best fit value is $N_{\rm H_{int}} = 4 \times 10^{19}$ cm$^{-2}$
and the 90\% upper limit is $2 \times 10^{20}$ cm$^{-2}$.  This will
be discussed in terms of a reddening correction to the optical data in
\S 4.1.

\subsubsection{Jet Spectrum}\label{sec:Jet}

With so few photons in the jet, no attempt to fit a spectrum
was made. However if the events are binned into three broad channels
($<$0.5 keV, 0.5-2.5 keV, and $>$2.5 keV), then there are 3, 18, and 5
events in these bands.  When this is compared with a similar
breakdown for the core, we conclude that within the statistical errors
of the jet counts, there is no evidence for its spectrum being
significantly different (particularly harder or softer) than that of
the core.

\begin{figure}
\plotone{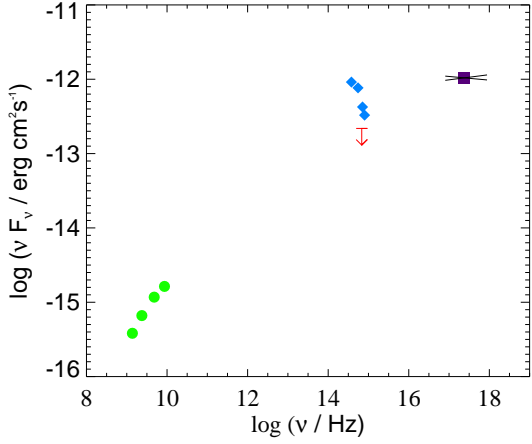}
\caption{\footnotesize The multi-wavelength spectral energy
distribution of the AGN shows the X-ray emission to be anomalously bright as
compared with
that at lower frequencies.  The arrow shows the upper limit given by \citet{Wetal}. The 1 $\sigma$ errors are smaller than the points plotted.\label{fig:multispectrum}}
\end{figure}
\begin{figure}
\plotone{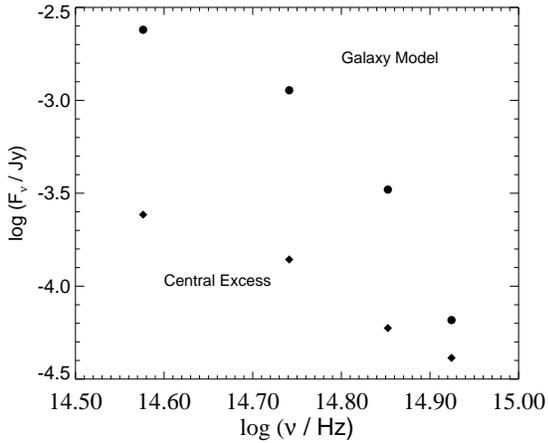}
\caption{\footnotesize The total flux measured within apertures of
the size of the band PSF and 9.66$''$ respectively, for the excess
emission and the galaxy model.  The 1 $\sigma$ errors are smaller than the points plotted.\label{fig:optical}}
\end{figure}

\section{Discussion}\label{sec:discussion}

Our NTT optical observations have detected a central excess emission.
Figure \ref{fig:multispectrum} shows that this emission has a steep
spectrum compared with the flatter AGN emission seen at other wavelengths.  However, the
spectrum of the excess is blue compared with the
emission from the elliptical galaxy (Figure \ref{fig:optical}), as
expected for an AGN.
One possibility is that the host galaxy is more cuspy than a simple elliptical. This is unaccounted for by the model we have fit and so the excess is contaminated with some level of starlight.
It is of course possible that the extrapolation of the galaxy profile
to small radii does not provide a sufficiently accurate baseline upon
which to measure the AGN optical light.   This might be supported by
the fact that the upper limit of 32$\mu$Jy in the blue estimated by
\citet{Wetal} is a factor of $\sim 4$ below the NTT excess (Table~2).
However, the discrepancy between these values can equally well be
explained by variability.  The radio is measured to vary by $\sim
20\%$, and larger variability is expected from high-energy electrons
emitting in the optical under a synchrotron-radiation interpretation
(see below).  In either case there are difficulties in developing a
simple interpretation of the source in terms of known emission
mechanisms, as discussed below.

\begin{figure}
\plotone{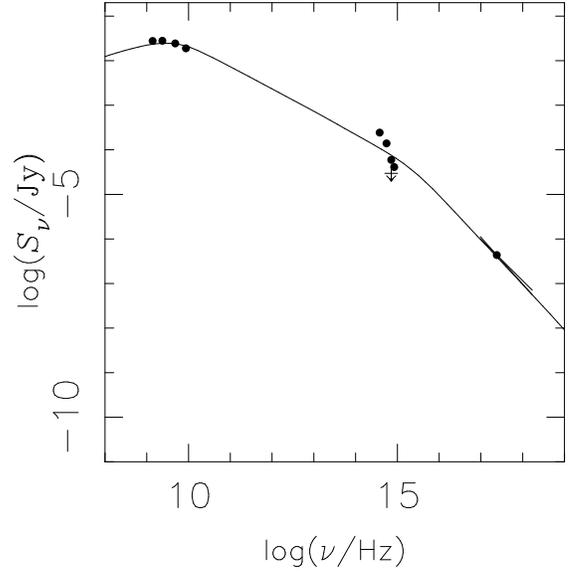}
\caption{\footnotesize The core spectral distribution shown with a
synchrotron model where an electron spectrum with $N_\gamma \propto
\gamma^{-2}$ breaks to $N_\gamma \propto \gamma^{-3}$.  Such a simple
model can explain the radio and X-ray fluxes and spectra.  However,
the optical is overpredicted in the blue and has the wrong spectrum, and a
spectral break above the optical bands is extreme, as such breaks in
other low-power radio galaxies and BL~Lac objects are in the infrared.\label{fig:break}}
\end{figure}

As known previously, the low optical intensity and lack of nuclear
emission lines means that the optical emission is not as expected from
an optically-thick accretion disk associated with a supermassive black
hole radiating at close to the Eddington luminosity.  We therefore
associate the nuclear optical emission with synchrotron emission from
an unresolved jet, as is typical for FRI radio galaxies and BL Lac
objects \citep{hw2000,ccc99}.  The difficulty
then occurs in explaining the X-ray emission whose spectrum is now
well measured with {\it Chandra}.  An extrapolation of the X-ray
spectrum to low energy falls well above the radio, but the radio and
X-ray emission could be described by a broken power law, as is typical
in other sources \citep[e.g.,][]{hbw2001}, albeit that the break frequency
is at a higher frequency than is typical.  Such an example model
spectrum is shown in Figure~\ref{fig:break}.  However, the optical is
a problem for such a model.  If we say that the spectrum of the excess
is that of the AGN, and that intensity variability explains the
inconsistency with the previous upper limit, then we have to explain
why the optical spectrum is much steeper than expected for the model.
The obvious thing to consider is additional dust reddening that has
not been taken into account.  The relationship between extinction and
column density of atomic hydrogen is given by \citet{Betal} as

$$E_{\rm B-V} = ~{\rm max}~[0, (-0.055 + 1.987 \times 10^{-22} N_{\rm
H})]$$

\noindent
(see e.g., \citet{Wilkes}).  In order to flatten the spectrum of the
optical excess to agree with the single component synchrotron model in
Figure \ref{fig:break} we would need $N_{\rm H}$ of
$1.5\times10^{21}\rm{cm}^{-2}$.  However, we have placed a 90\% upper
limit on the level of intrinsic $N_{\rm H}$ of
$2\times10^{20}\rm{cm}^{-2}$ based on the X-ray data. Therefore
reaching consistency would point either to a highly anomalous
gas-to-dust ratio, i.e., dust unlike Galactic dust, or that reddening
cannot be the explanation for the steep optical spectrum.  The only
way to bring things into consistency with a simple broken-power-law
synchrotron model would be to suggest that variability and
contamination of starlight in the optical excess conspire in such a
way that the underlying AGN spectral index matches the expected value,
and that the AGN was brighter in 2004 than in 1996.  We cannot rule
out such a conspiracy.

\begin{figure}
\plotone{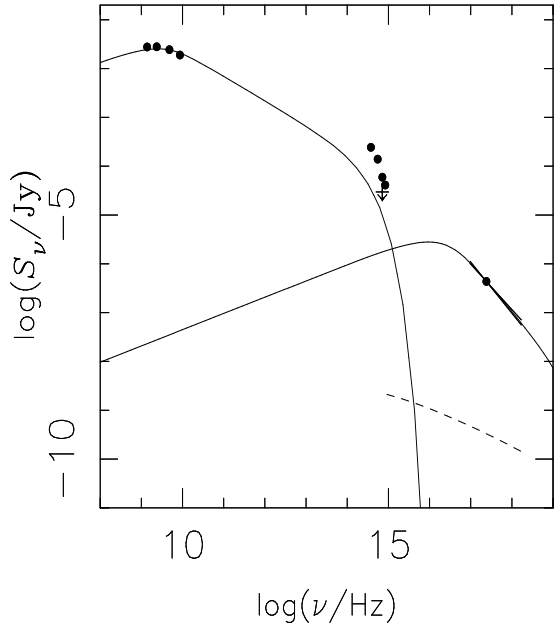}
\caption{\footnotesize  The
low-energy synchrotron component is modelled from an electron number
spectrum with $N_\gamma \propto \gamma^{-2}$, where $\gamma$ is
electron Lorentz factor, and $\gamma_{\rm min}=120$, $\gamma_{\rm
max}=4 \times 10^4$, in a minimum-energy magnetic field of strength
8~$\mu$T (80 mG) and a source of radius 0.2 pc.
The dashed line is the SSC prediction.  No
relativistic beaming is assumed (this would cause the dashed line to
drop a little).  The higher energy model is
an extra high-energy synchrotron component with
an extraordinarily high $\gamma_{\min}$ of $4 \times 10^5$ (assuming
a minimum-energy magnetic field). \label{fig:twosynch}}
\end{figure}

Other possible explanations for the X-ray emission can be considered,
such as inverse Compton scattering.  There is no evidence for an
intense photon field from a radiatively-efficient accretion disk, and,
since the X-ray emission is from within the nucleus, scattering of
cosmic microwave background photons gives a low yield, even if
assisted by relativistic boosting.  The yield from synchrotron self
Compton is maximized if the source is made as compact as possible.  We
adopt a sphere of radius 0.2~pc, that was argued by \citet{Wetal} to be the
smallest size reasonable to prevent the radio emission from being
entirely self absorbed.  The level of synchrotron self-Compton X-ray
emission is insensitive to the high-energy cut-off of the synchrotron
radiation, and we have calculated a result under a minimum-energy
magnetic field for a synchrotron spectrum that cuts off rather than
overproduces the optical excess.  The result is shown as the dashed
line in Figure~\ref{fig:twosynch}.  The X-ray intensity is
underpredicted by more than two orders of magnitude.  The intensity
can be increased by reducing the magnetic field below the
minimum-energy value, but even then the spectral slope will not match
observations.  We therefore rule out inverse Compton emission as the
origin of the X-rays.

The usual alternative is to add a second high-energy
component of synchrotron radiation, and that is shown as the
high-energy line in Figure~\ref{fig:twosynch}.
This has been suggested as the explanation of the X-ray emission in
resolved regions of some quasar jets \citep{rmn2000,jester}.
The difficulty is that in order not to overpredict the optical, the
low-energy electron cut-off in this second component must be at very
high energy (the Lorentz factor is  $4 \times 10^5$ in the example
in Figure~\ref{fig:twosynch}).
There is a difficulty understanding
how this second component
can be produced, such that there are no low-energy electrons.

\section{Summary}

Our new optical, X-ray and radio observations of J2310-437 confirm
it as an AGN with no more than a weak optical continuum, a relatively
weak radio core and jet, but a strong unabsorbed X-ray power-law
continuum.  It is at the extreme of such a class of objects, and
thus challenges an understanding in terms of the emission mechanisms
applied most commonly to related active galaxies.
Specifically, the X-ray emission is difficult to understand.  An origin
as the corona of a geometrically-thin radiatively-efficient accretion
disk is ruled out by the weak optical emission.  The X-ray intensity
and spectrum appear to rule out an inverse Compton origin from a
source component of plausible size, although in an attempt to explore
this aspect further we are making VLBI observations to examine the
structure and orientation of the jet on the smallest spatial scales on
which the core is resolved.
An X-ray synchrotron origin appears to be most likely, and under this
model our optical observations of a central excess that may wholly or
in part be associated with the AGN place constraints on the
lower-energy population of electrons.  To help tie down the level of
association of the optical excess with AGN synchrotron emission,
upcoming photometry with Spitzer will test if the longer-wavelength IR
bands are dominated by non-thermal AGN emission, and test the
extension of the spectrum of the measured central optical excess into
the IR.  The longer IR wavelengths will allow us to probe for
anomalous dust properties that might be affecting the optical
spectrum.  For a single population of electrons that extend down to
radio-emitting energies to be present, the optical emission must have
been variable by factors of at least 4 between the different epochs of
our observations.  A better optical separation from the bright galaxy
light, as would be possible with HST, over multiple epochs of
observation, would test this possibility.
If the tests fail, we seem forced to explain the X-ray emission as
synchrotron radiation from a population of electrons with a very high
low-energy cut-off --- a Lorentz factor $\sim 4 \times 10^5$ in our
nominal model, although component size and any level of departure from
minimum-energy will affect this value.  It remains open as to how such
an electron component might arise.

\acknowledgments 
AFB acknowledges support from an UK Science and
Technology Facilities Council (STFC) studentship. We also acknowledge support from NASA (NAS8-03060).

\bibliographystyle{astron}
\bibliography{refs}

\clearpage

\end{document}